\documentclass[12pt]{elsart}
\usepackage{graphicx}
\begin{document}
\begin{frontmatter}
\title{\boldmath Characterization of the electronic properties of YB$_4$ and
YB$_6$ using $^{11}$B NMR and first-principles calculations}

\author{B. J\"ager$^a$, S. Paluch$^b$, W. Wolf$\,{}^c$, P. Herzig$^a$, O.~J. \.Zoga{\l}$^b$,}
\author{N. Shitsevalova$^d$, Y. Paderno$^d$}
\author{}
\address{$^a$Institut f\"ur Physikalische Chemie, Universit\"at Wien,
W\"ahringer Stra{\ss}e 42, 1090~Vienna, Austria\\
$^b$Institute for Low Temperature and Structure Research, Polish Academy of Sciences,
P.~O. Box 1410, 50-950 Wroc{\l}aw, Poland\\
$^c$Materials Design s.~a.~r.~l., 44, av. F.-A. Bartholdi, 72000 Le Mans, France\\
$^d$Institute for Problems of Materials Science, Academy of Sciences of Ukraine,\\
3 Krzhyzhanovsky str., 03680 Kiev, Ukraine}
\corauth[corr]{Corresponding author. Fax: +43-1-4277-9524.}
\ead{peter.herzig@univie.ac.at (P. Herzig)}
\clearpage
\begin{abstract}
Two compounds, tetragonal YB$_4$ and cubic YB$_6$, have been investigated
by electric-field gradient (EFG) and Knight shift measurements
at the boron sites using the $^{11}$B nuclear magnetic resonance (NMR) technique and by performing
first-principles calculations. In YB$_6$ $^{11}$B ($I=3/2$) NMR spectra
reveal patterns typical for an axially symmetric field gradient with a quadrupole
coupling frequency of $\nu_Q=600\pm15$~kHz. In the second boride (YB$_4$)
three different EFGs were observed corresponding to the three
inequivalent crystallographic sites for the boron atoms ($4h$, $4e$, and $8j$).
They correspond to: $\nu_Q(4h)=700\pm30$~kHz with an asymmetry parameter
$\eta=0.02\pm0.02$, $\nu_Q(4e)=515\pm30$~kHz, $\eta=0.00+0.02/-0.00$, and
$\nu_Q(8j)=515\pm40$~kHz, $\eta=0.46\pm0.08$.  The Knight shifts measured by
Magic-Angle Spinning (MAS) NMR
at room temperature are very small being $0.6\pm8$~ppm and $-1\pm8$~ppm for YB$_4$
and YB$_6$, respectively. For the theoretical calculations
structure optimizations were performed as a first step. For the obtained structural
parameters the EFGs were computed within the local-density
approximation. Very satisfactory agreement between experimental and theoretical
results is obtained both for the structural parameters and the B EFGs thus confirming
the underlying structural models. In addition to the EFGs, band structures,
densities of states, and valence-electron densities are presented and the bonding situation
in the two yttrium borides is discussed. The band-structure results are compatible
with the very low values for the Knight shifts mentioned above.
\end{abstract}

\begin{keyword}
Metal borides; Electronic band structure; Electric field gradient; Chemical bonding;
NMR
\end{keyword}

\end{frontmatter}
\clearpage

\section{Introduction}
Through their specific properties, transition metal borides
became attractive both from scientific and technical point of view. For
instance, YB$_6$, like some other rare-earth and actinide hexaborides, 
shows superconductivity below 7.1~K~\cite{mga-s-530-68} and has been considered
as a possible candidate for high-temperature thermoelectric materials~\cite{imu-i-721-01}.

Detailed knowledge of the electronic structure of a material is a key ingredient for
an in-depth understanding of many of its macroscopic features. A combination of
experimental methods as well as computations on an \textit{ab-initio\/} level has proved very
efficient, in particular relating electronic structure calculations to results of
nuclear magnetic resonance (NMR) technique. The value of the electric field gradient 
(EFG), measured by NMR quadrupole interaction, is directly determined by the charge 
distribution around the nucleus. Thus, theoretical EFG studies based on the electronic
structure are important in order to give a reliable interpretation of the experimental
data. One of the strengths of NMR measurements is that, being a microscopic tool,
they are sensitive to the symmetry of the crystallographic sites and in particular to
the electron density in the vicinity of the nucleus. In a recent study, this sensitivity 
was used to provide criteria for the determination of crystal structures that are
still under debate~\cite{zwh-prb-214110-01,wh-prb-224112-02,wh-jac-print-03}.

In the present paper, we report the EFG values for YB$_6$ and YB$_4$ and the determination of
the Knight shifts by using the Magic-Angle Spinning (MAS) technique. The
experimental EFGs are compared and interpreted with the theoretical ones obtained from 
\textit{ab-initio\/} calculations. 

The structure of cubic YB$_6$ (space group $Pm\overline{3}m$, no. 221) consists of a simple
cubic lattice of corner-connected B$_6$ octahedra with Y atoms filling the cubic
holes. In this and other borides with the same structure the B--B distance between
neighbouring octahedra is slightly shorter than within the octahedron (for YB$_6$:
1.64~\AA\ compared to 1.75~\AA). In agreement with this crystallographic structure there 
is only a single EFG value. A previous $^{11}$B NMR study for YB$_6$~\cite{okk-cjp-787-96}
reported the
spin-lattice relaxation time and the spectrum. However, the authors were unable to
determine the EFG because of the widely broadened satellite lines and the weak NMR
signal used in their instrumentation.

On the other hand, tetragonal YB$_4$ (space group $P4/mbm$, no. 127) has a quasilayer
structure with alternating sheets of Y atoms and B$_6$ octahedra linked together
laterally by B$_2$ units. In this structure five different nearest-neighbour B--B
distances between 1.64~\AA\ and 1.81~\AA\ are observed. In YB$_4$ the boron atoms are
situated in three different crystallographic positions with different EFGs for each
of them, as will be demonstrated in the present paper.

\section{Structure optimization}
For confirmation of the validity of the available experimental lattice parameters
and the atomic positions,
calculations have been performed using the Vienna \textit{ab-initio\/} simulation package
(VASP)~\cite{vasp1,kf-prb-11169-96,kf-cms-15-96}. By this method the Kohn--Sham
equations of density-functional theory~\cite{hk-pr-b864-64,ks-pr-a1133-65} with
periodic boundary conditions are solved within a plane-wave basis set with electron--ion
interactions described by the projector augmented wave (PAW)
method~\cite{b-prb-17953-94,kj-prb-1758-98}. For exchange and correlation the general
gradient approximation (GGA)~\cite{pcv-prb-6671-92} was applied.
The structural parameters were calculated by atomic forces and stress-tensor minimization.
For each of the two borides two different energy cutoffs for the plane-wave basis
were used (400~eV and a cutoff higher by at least a factor of 2), which lead to
practically the same results, so that adequate convergence
is ensured. The experimental and calculated structural parameters for YB$_4$ and 
YB$_6$ are given in Tables~1 and 2, respectively. For cubic YB$_6$, where the lattice
parameter $a$ and the positional parameter of the B atom are the only free parameters,
the minimum-energy structure was also
found by the full-potential linearized augmented plane-wave
(FLAPW) method within the LDA approximation which is described in the next Section. 

The structural parameters optimized by VASP have been used to calculate
the EFGs for YB$_6$ and YB$_4$.

\begin{table}
\caption{Experimental and calculated structural parameters for YB$_4$
adopting the ThB$_4$ structure ($P4/mbm$, no.~127). The lattice
parameters are in \AA.}
\begin{tabular}{lllll}
\hline
Lattice      &Positional   &                 \\
parameters   &parameters   &Atom&Site&Remarks\\
\hline
$a=7.111$    &$x=0.3179$   &Y   &$4g$&exp., Ref.~\cite{gve-cras-145-80}\\
$c=4.017$    &$z=0.2027$   &B(1)&$4e$&       \\
             &$x=0.0871$   &B(2)&$4h$&       \\
             &$x=0.1757$   &B(3)&$8j$&       \\
             &$y=0.0389$   &    &    &       \\[5pt]
$a=7.1035$   &             &    &    &exp., Ref.~\cite{kr-hthp-453-90}\\
$c=4.0206$   &             &    &    &       \\[5pt]
$a=7.1091$   &$x=0.3182$   &Y   &$4g$&calc., VASP (GGA)\\
$c=4.0280$   &$z=0.2030$   &B(1)&$4e$&       \\
             &$x=0.0870$   &B(2)&$4h$&       \\
             &$x=0.1760$   &B(3)&$8j$&       \\
             &$y=0.0386$   &    &    &       \\
\hline
 \end{tabular}
 \label{table1}
 \end{table}

\begin{table}
\caption{Experimental and calculated structural parameters for YB$_6$
adopting the CaB$_6$ structure ($Pm\overline{3}m$, no.~221). The lattice
parameters are in \AA, the positional parameter refers to site $6f$ of B.}
\begin{tabular}{lll}
\hline
Lattice      &Positional   &       \\
parameter $a$&parameter $x$&Remarks\\
\hline
4.102        &             &exp., Ref.~\cite{zbm-spc-723-71}\\
4.1025       &0.202        &exp., Ref.~\cite{b-rr-52-77}\\
4.113        &             &exp., Ref.~\cite{k-jjap-15-94}\\
             &0.199        &exp., Ref.~\cite{too-jpsj-2304-99}\\
4.0436       &0.1989       &calc., FLAPW (LDA)\\
4.1004       &0.1988       &calc., VASP (GGA)\\
\hline
 \end{tabular}
 \label{table2}
 \end{table}

\section{Electric-field gradients}
\subsection{Experimental}
To avoid skin-depth effects for better RF penetration, the samples
were used in powder form and obtained by crushing the single crystals.
The crystals were obtained by crucible free inductive zone melting of
the YB$_4$ and the YB$_6$ source rods that provides a high purity and
single-phase material. They, in turn, were prepared in the following steps:
\begin{enumerate}
\item Mixing of Y$_2$O$_3$ (purity 99.999\%) and boron components in proper ratios.
\item Synthesis of the YB$_4$ and the YB$_6$ powders by routine solid-state
reaction of borothermic reduction:
\begin{displaymath}
\begin{array}{c}
\mbox{Y}_2\mbox{O}_3+11\mbox{B} \rightarrow 2\mbox{YB}_4+3\mbox{BO}\uparrow\\
\mbox{Y}_2\mbox{O}_3+15\mbox{B} \rightarrow 2\mbox{YB}_6+3\mbox{BO}\uparrow
\end{array}
\end{displaymath}
\item Slip casting of source rods, removal of the binder and sintering of the
rods in the vacuum.
\end{enumerate}
The $^{11}$B NMR experiments were carried out with a Bruker DSX 
Avance spectrometer at a frequency of 96.29~MHz. The static spectra were 
obtained by the Fourier transform of the free induction decay (FID) 
following a short single pulse (in the range of 0.9--2.5~$\mu$s). The spectra 
contain up to 256 accumulations with a repetition time of 20~s. A 4~mm MAS
probe was used with rotation frequencies between 7 and 9~kHz. The chemical
shifts are given with respect to external BF$_3$Et$_2$O. In Fig.~1 the
$^{11}$B NMR spectra for YB$_4$ and YB$_6$ are given. 
\linespread{1.4}
\begin{figure}[p]
\begin{center}
\vspace*{-1cm}
\includegraphics[width=0.85\hsize]{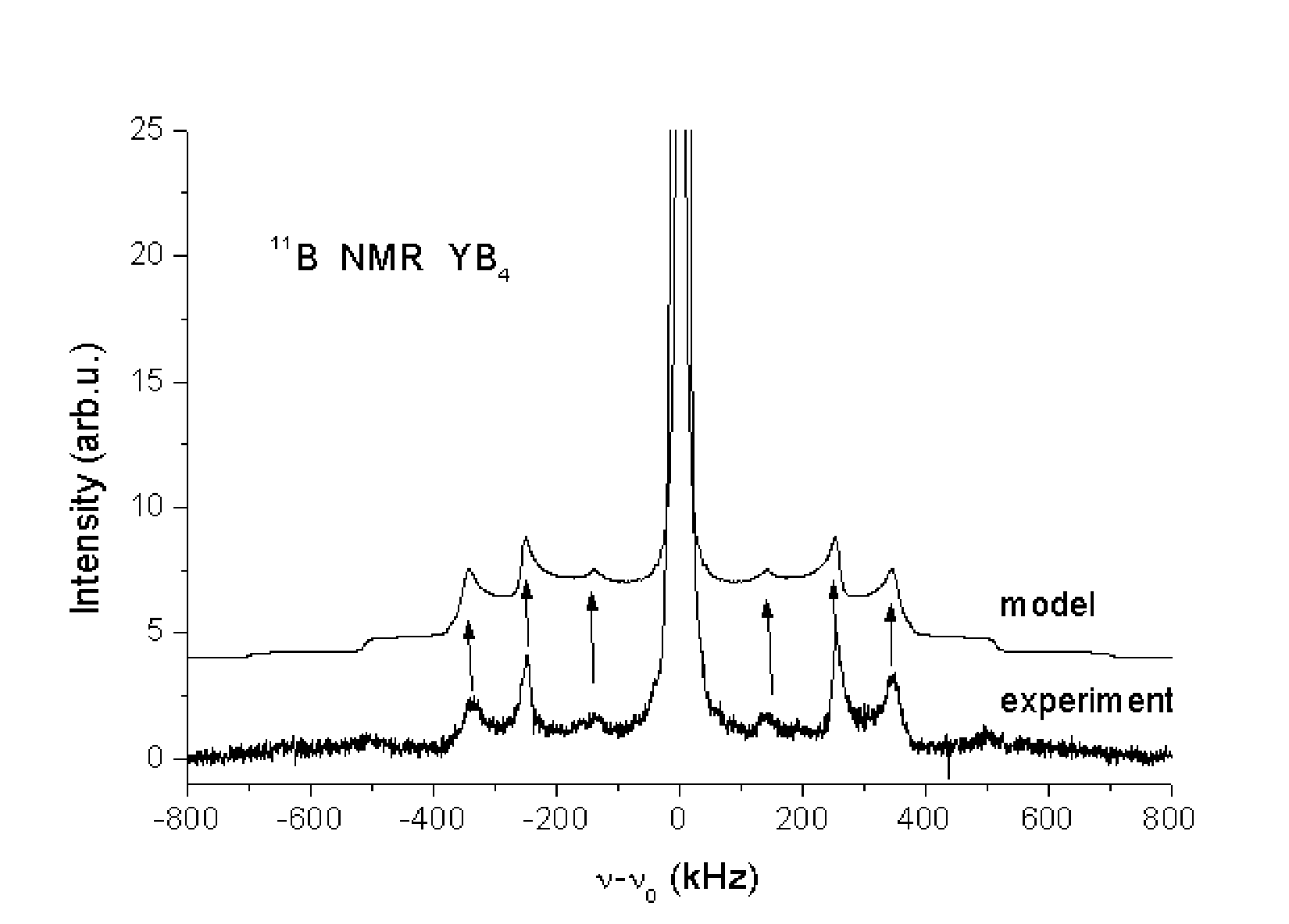}\\
\includegraphics[width=0.85\hsize]{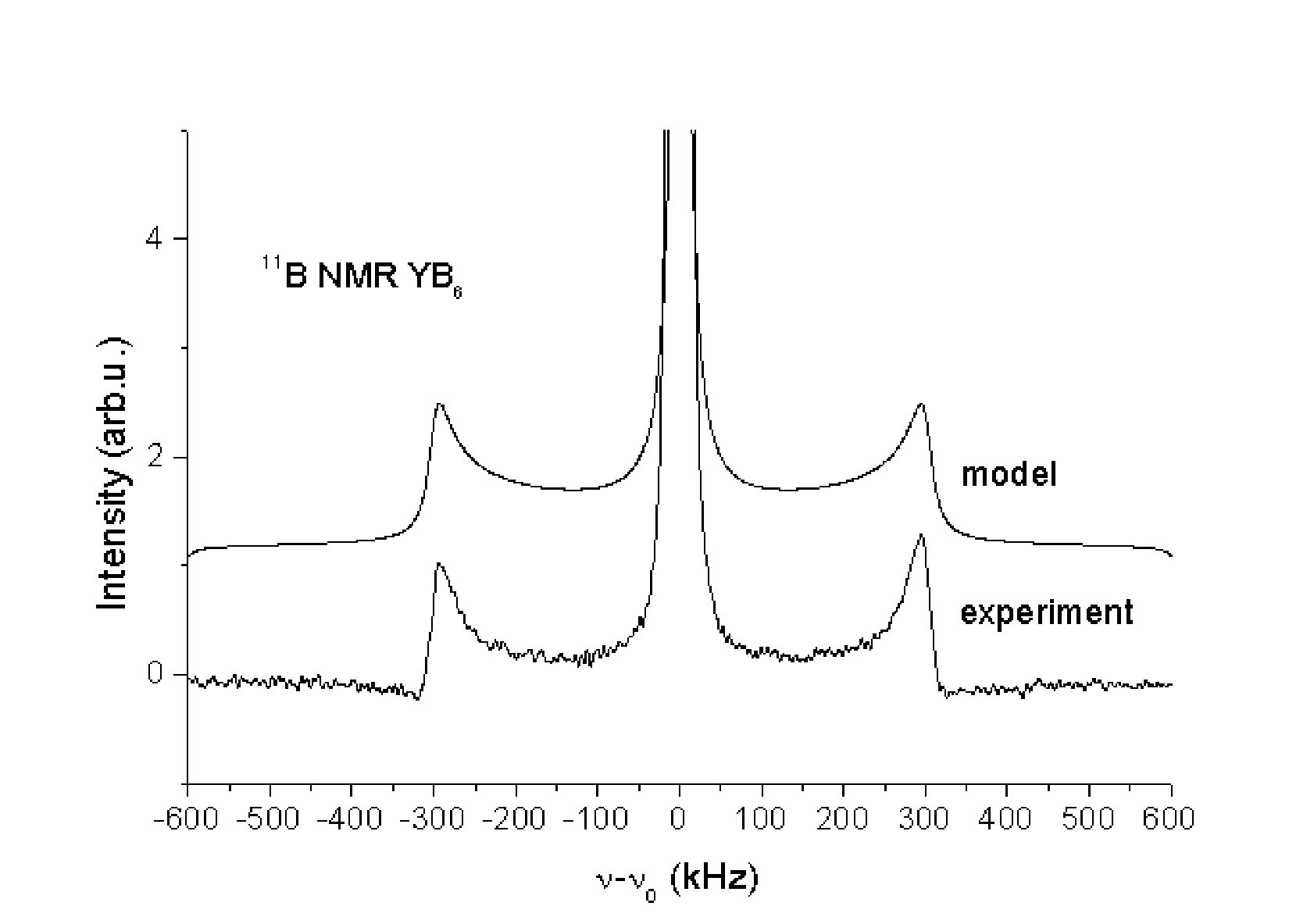}
\end{center}
\caption{Experimental and theoretical (model) powder-pattern $^{11}$B spectrum for
YB$_4$ (top) and YB$_6$ (bottom). For better readability the model spectra are
shifted up with respect to the experimental ones. For YB$_4$ the arrows indicate positions
of $90^\circ$ singularities ($\eta\sim0$) for B(2) (outermost arrows) and
B(1) (middle arrows) except for the singularity for B(3) (innermost
arrows) where the angle is different from $90^\circ$ ($\eta=0.46$).}
\end{figure}

\subsection{First-principles calculations}
The all-electron band-structure calculations for the calculation of EFGs
are based on the density-functional
theory~\cite{hk-pr-b864-64,ks-pr-a1133-65} (DFT) and the local-density
approximation and have been performed by the linearized augmented plane-wave
(LAPW) method~\cite{a-prb-3060-75} in its full-potential
version~\cite{ka-jpf-2041-75,wkw-prb-864-81,jf-prb-561-84,moj-jmmm-1091-86}
(FLAPW) using an exchange-correlation potential by Hedin and
Lundqvist~\cite{hl-jpc-2064-71,hl-jpp-c3-73-72}.

The EFGs have been calculated from the $l=2$ components of the Coulomb
potential near the nuclei. The formalism by Herzig~\cite{h-tca-323-85} and
Blaha et al.~\cite{bsh-prl-1192-85} has also been employed to split the calculated EFG
components into the contributions from the surrounding electrons within the
respective muffin-tin sphere (``sphere contribution'') and the remainder
that comes from outside this sphere (``lattice contribution''). This
partitioning depends, to a small extent, on the choice of the muffin-tin
radii. The valence contribution can be split further into the allowed $ll'$
contributions (only $sd$, $pp$, $pf$ are important in the present context)
which provide useful information about the influence of particular $l$-like
wave functions on the EFGs~\cite{bsd-prb-2792-88}. As is common practice
the EFG component with the largest absolute value is always designated as
$V_{zz}$. For further algorithmic details on the EFG calculation
see our recent paper~\cite{wh-prb-224112-02}.

\subsection{EFGs: Results and discussion}
Figure~1 shows the $^{11}$B spectra which are typical for a nuclear spin
$I=3/2$ in the presence of first-order quadrupole effects. For YB$_6$ (bottom of Fig.~1)
the separation of the satellite lines is given by $\delta\nu=\nu_Q\,(3\cos^2\Theta-1)$,
where $\Theta$ are the powder singularities. $\Theta=90^\circ$ thus yields $\nu_Q=e^2qQ/2h$.
Here, $eq=V_{zz}$ and $Q$ is the largest component of the electric-field-gradient tensor
and the nuclear quadrupole moment, respectively. The axially symmetric field gradient is
consistent with the local symmetry at the boron site. The existence of three inequivalent
boron sites in YB$_4$ is the origin of the more complex satellite spectrum compared to YB$_6$
(Fig.~1). The $\nu_Q$ value and the asymmetry parameter $\eta=(V_{xx}-V_{yy})/V_{zz}$ were obtained
using the \textit{dmfit\/} simulation program~\cite{mfc-mrc-70-02}. The $\nu_Q$ value translates
into $C_q=2\nu_Q$ and it follows for $V_{zz}=4.136\times10^{19}\,C_q/Q$ (in $\mbox{V}/\mbox{m}^2$)
when the appropriate $Q$ value (in barns) for $^{11}$B is used and $C_q$ is inserted
in units of MHz. Since different nuclear
quadrupole moments can be found in the literature, the experimental values for $V_{zz}$ were
determined for two different values, namely $Q=0.04$~b and $Q=0.0355$~b~\cite{nw-nmr-1-78}.
The former $Q$ value has been taken from the almanac of the spectrometer manufacturer
Bruker and is very close to $Q=0.04059$~\cite{so-jcp-5051-91} recommended
by Pyykk\"o~\cite{p-zn-189-91} in his compilation of nuclear quadrupole moments. 
The results are shown in Tables~3 and 4 for YB$_6$ and YB$_4$, respectively. Experimentally
only the absolute value of $V_{zz}$ and $\eta$ can be determined for the powder specimens.

\begin{table}
  \caption{Calculated B EFGs (in $10^{20}$ V/m$^2$) for YB$_6$ for the
           optimized structure compared to the experimental results assuming an
           $^{11}$B nuclear quadrupole
           moment of 0.04 and 0.0355~$|e|\times10^{-28}$~m$^2$, respectively.}
  \begin{tabular}{cp{0.3pc}cp{0.3pc}c}
  \hline
Calc.       &&Exp.         &&Exp.\\
Opt. struct.&&$Q=0.04$     &&$Q=0.0355$\\
\cline{1-1}\cline{3-3}\cline{5-5}
$V_{zz}$    &&$|V_{zz}|$   &&$|V_{zz}|$\\[2pt]
  \hline
$-13.5$     &&$ 12.4\pm0.3$&&$ 14.0\pm0.3$\\[2pt]
  \hline
  \end{tabular}
 \label{table3}
\end{table}

\begin{table}
  \caption{Calculated B EFGs (in $10^{20}$ V/m$^2$) for YB$_4$ for the
           experimental structure~\cite{gve-cras-145-80} and the optimized structure
           compared to the experimental results assuming an $^{11}$B nuclear quadrupole
           moment of 0.04 and 0.0355~$|e|\times10^{-28}$~m$^2$, respectively.}
  \begin{tabular}{@{}lcrl@{}p{0.0pc}rl@{}p{0.0pc}rl@{}p{0.0pc}rl@{}}
  \hline
    &&\multicolumn{2}{c}{Calc.}&&\multicolumn{2}{c}{Calc.}&&
      \multicolumn{2}{c}{Exp.}&&\multicolumn{2}{c}{Exp.}\\
    &&\multicolumn{2}{c}{Exp. struct.}&&\multicolumn{2}{c}{Opt. struct.}&&
      \multicolumn{2}{c}{$Q=0.04$}&&\multicolumn{2}{c}{$Q=0.0355$}\\
      \cline{3-4}\cline{6-7}\cline{9-10}\cline{12-13}
Site&$i$&\multicolumn{1}{c}{$V_{ii}$}&\multicolumn{1}{c}{$\eta$}&&
        \multicolumn{1}{c}{$V_{ii}$}&\multicolumn{1}{c}{$\eta$}&&
        \multicolumn{1}{c}{$|V_{ii}|$}&\multicolumn{1}{c}{$\eta$}&&
        \multicolumn{1}{c}{$|V_{ii}|$}&\multicolumn{1}{c}{$\eta$}\\[2pt]
  \hline
B(1)&$z$&$-11.0$&$0.0 $&&$-10.8$&$0.0 $&&$10.6\pm0.6$&$0.0        $&&$12.0\pm0.6$&$0.0 $\\[2pt]
B(2)&$x$&$ -8.1$&$    $&&$ -8.2$&$    $&&$          $&$           $&&$          $&$    $\\
    &$y$&$ -8.3$&$    $&&$ -8.4$&$    $&&$          $&$           $&&$          $&$    $\\
    &$z$&$ 16.4$&$0.02$&&$ 16.6$&$0.01$&&$14.5\pm0.6$&$0.02       $&&$16.3\pm0.6$&$0.02$\\[2pt]
B(3)&$x$&$  2.7$&$    $&&$  2.9$&$    $&&$          $&$           $&&$          $&$    $\\
    &$y$&$  7.9$&$    $&&$  7.9$&$    $&&$          $&$           $&&$          $&$    $\\
    &$z$&$-10.6$&$0.48$&&$-10.8$&$0.47$&&$10.6\pm0.9$&$0.46\pm0.09$&&$12.0\pm0.9$&$0.46\pm0.08$\\
  \hline
  \end{tabular}
 \label{table4}
\end{table}

For YB$_6$ the B EFG corresponds to the axially symmetric case (site symmetry $C_{4v}$)
where there is only one independent EFG component. In YB$_4$ the situation is more
complicated. The three crystallographically distinct B sites (site symmetries: $C_4$ for
B(1), $C_{2v}$ for B(2), and $C_s$ for B(3)) lead to one, two, and three independent
EFG components, respectively. This means that for B(2) and B(3) a non-zero asymmetry
parameter $\eta$ is observed.

The comparison between calculated and measured EFGs shows very good agreement,
if the absolute values are considered. In general for the EFG tensor
either one or two components are negative. The corresponding principal axes
are determined by the strongest bonding interactions. In the case of YB$_4$
the negative EFG component for B(1) ($V_{zz}$) is related to the axis of
the inter\-octahedral B--B bond along the $c$ axis, for B(2) the two negative components
($V_{xx}$ and $V_{yy}$) refer to the (001) plane through B(2) and its 
three B neighbours, and finally for B(3) the negative EFG component ($V_{zz}$)
refers to the axis of the bond between the adjacent atoms of a B$_6$ octahedron
and a B$_2$ unit.

Now the splits of the B EFGs into the lattice and sphere contributions and the
latter into their main components ($sd$, $pp$, and $pf$) are considered (Tables~5
and 6). The sphere contribution is the one with the largest absolute value,
but also the lattice contribution is relatively large. This behaviour has been
observed previously for borides and other second-group elements by 
Schwarz~\etal~\cite{srb-zn-527-96}. For the H atom in hydrides the lattice
contribution is even larger than the sphere contribution, as has been found
recently~\cite{wh-prb-224112-02}. It is not surprising that the $pp$ component
dominates the sphere contribution with the $sd$ and $pf$ components being the
largest other components.
\begin{table}
\caption{Split of the B EFG for YB$_6$ (optimized structure) into lattice and
sphere components and the latter into its main contributions, i.~e., $sd$, $pp$, and $pf$.
All $V_{zz}$ values are in units of $10^{20}$ V/m$^2$.}
\begin{tabular}{lrrrrrr}
\hline
Site&\multicolumn{1}{c}{$V_{zz}$}&\multicolumn{1}{c}{$V_{zz}^{lat}$}&
                \multicolumn{1}{c}{$V_{zz}^{sph}$}&\multicolumn{1}{c}{$V_{zz}^{sd}$}&
                \multicolumn{1}{c}{$V_{zz}^{pp}$}&\multicolumn{1}{c}{$V_{zz}^{pf}$}\\
\hline
B&$-13.5$& $ 9.0$&$-22.5$&$ -1.9$&$-17.8$&$-2.3$\\
\hline
 \end{tabular}
 \label{table5}
\end{table}

\begin{table}
\caption{Split of the B EFGs for YB$_4$ (optimized structure) into lattice and
sphere components and the latter into its main contributions, i.~e., $sd$, $pp$, and $pf$.
All $V_{zz}$ values are in units of $10^{20}$ V/m$^2$.}
\begin{tabular}{lrrrrrr}
\hline
Site&\multicolumn{1}{c}{$V_{zz}$}&\multicolumn{1}{c}{$V_{zz}^{lat}$}&
                \multicolumn{1}{c}{$V_{zz}^{sph}$}&\multicolumn{1}{c}{$V_{zz}^{sd}$}&
                \multicolumn{1}{c}{$V_{zz}^{pp}$}&\multicolumn{1}{c}{$V_{zz}^{pf}$}\\
\hline
B(1)&$-10.8$& $ 7.2$&$-18.0$&$ -1.4$&$-14.6$&$-1.7$\\
B(2)&$ 16.6$& $-4.0$&$ 20.6$&$  1.3$&$ 17.8$&$ 1.3$\\
B(3)&$-10.8$& $ 5.0$&$-15.7$&$ -1.3$&$-12.6$&$-1.6$\\
\hline
 \end{tabular}
 \label{table6}
\end{table}
\subsection{Knight shifts}
The Knight shifts, $K$, measured for both compounds, are found to be very
small, i.~e., 0.6~ppm and $-1$~ppm for YB$_6$ and YB$_4$, respectively. The accuracy of
the $K$ values were estimated to be $\pm8$~ppm, where the uncertainty results from an
incomplete reduction of the strong dipolar interaction in the MAS experiment, mainly
between the boron nuclei. These values of $K$ are much smaller than those observed in
MgB$_2$, where $K(^{11}\mbox{B})=100$~ppm~\cite{gmv-prb-132506-02} and in YNi$_2$B$_2$C,
where $K(^{11}\mbox{B})\approx400$~ppm~\cite{sbt-prb-15341-96} (both values are for
room temperature). This comparison indicates that the values for the densities of
states (DOS) at the Fermi level for the boron sites are much lower than those in
the borides mentioned above. Especially the lack of $s$-type DOS, which is usually
the contribution with the largest hyperfine field, reduces significantly the shifts.
Indeed, the $s$-type DOS per boron atom is, in YB$_6$~\cite{som-cm-202015-02},
over 20 times smaller than in MgB$_2$~\cite{som-cm-202015-02}. Also the $p$-type DOS
per B atom is significantly smaller in YB$_6$ compared to MgB$_2$. These data are
compatible with the present electronic band structure calculations (see next Section)
which show that $s$- and $p$-type boron states are located mainly below the Fermi
energy and thus the contribution to the DOS at the $E_F$ is small. 

\section{Electronic structure and chemical bonding}
The first theoretical investigation of metal borides of general formula $M$B$_6$
has been performed fifty years ago by Longuet-Higgins and Roberts~\cite{lr-prsl-336-54}
using the tight-binding approximation. According to these authors the octahedral
arrangement of the B atoms leads to ten bonding states which,
when filled with the 18 B $s$ and $p$ plus two metal valence electrons,
will correspond to an insulator.
A metal will be obtained when the hexaboride of a trivalent metal is considered
instead of a divalent metal. In their paper the states below the Fermi level
are not only characterized in terms of their bonding or antibonding properties
but also as regards their participation in bonds within a single B$_6$ octahedron
or between neighbouring octahedra. Later~\cite{wem-prb-1859-77} the higher lying
valence and conduction
bands of YB$_6$ have been calculated by a discrete variational method in a
non-selfconsistent fashion. Recent electronic-structure calculations for YB$_6$
can be found in Ref.~\cite{imu-i-721-01} (DOS only) and in Ref.~\cite{som-cm-202015-02}
where also the band structure is given. For YB$_4$ only the DOS has been published
so far~\cite{imu-i-721-01,som-cm-202015-02}. 
In both references, however, the lowest B $2s$ band is not shown
although it is of crucial importance for the chemical bonding in both borides
as will be argued below. In the case of YB$_6$ Shein~\etal~\cite{som-cm-202015-02}
designate it erroneously as ``quasi-core B $2s$ band'' -- obviously mainly because
it is a low-lying band. In Fig.~2 the band structures for YB$_4$ and YB$_6$ are presented. 
\begin{figure}[p]
\begin{center}
\mbox{\includegraphics[width=0.5\hsize]{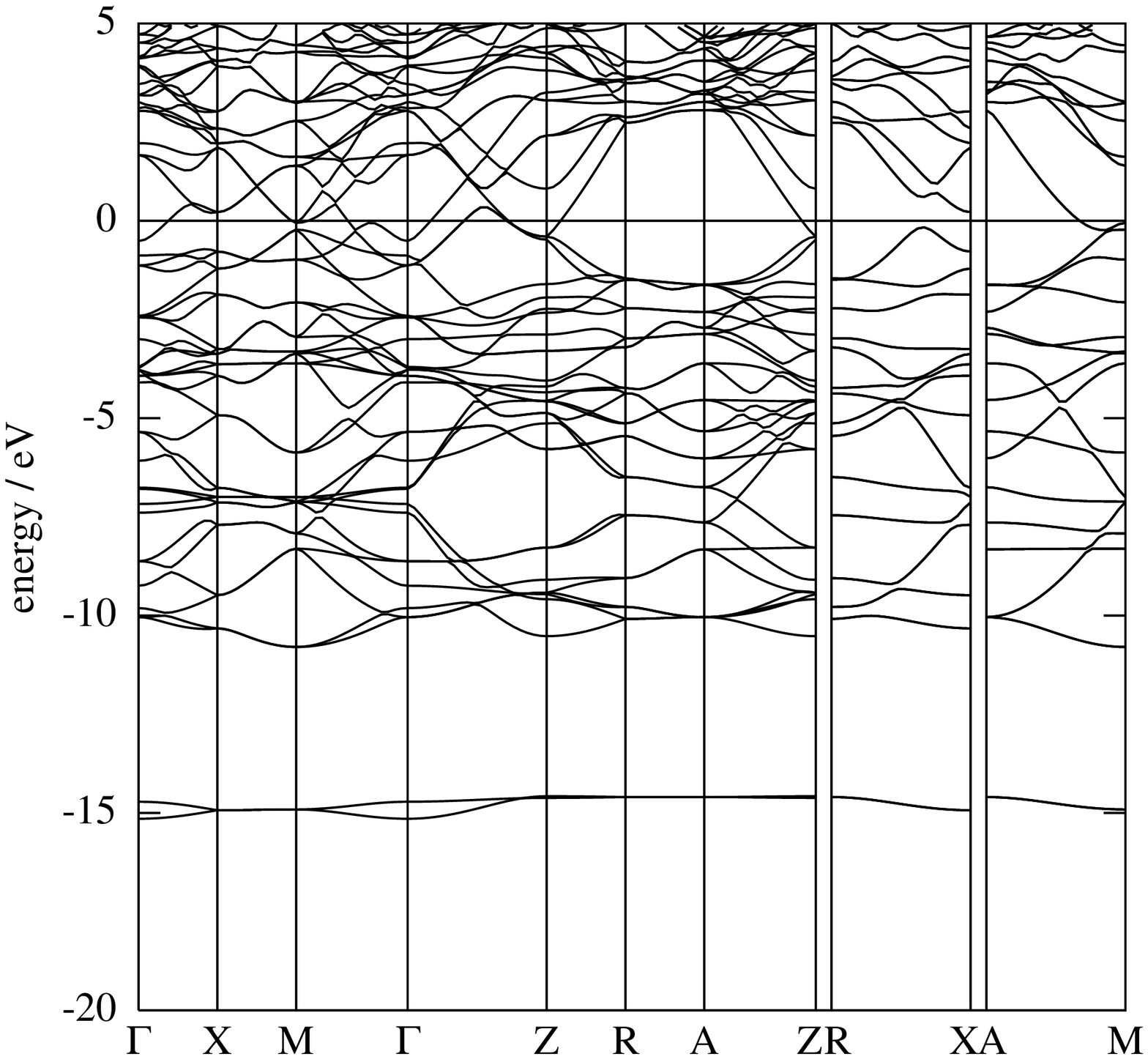}\hfil
\includegraphics[width=0.5\hsize]{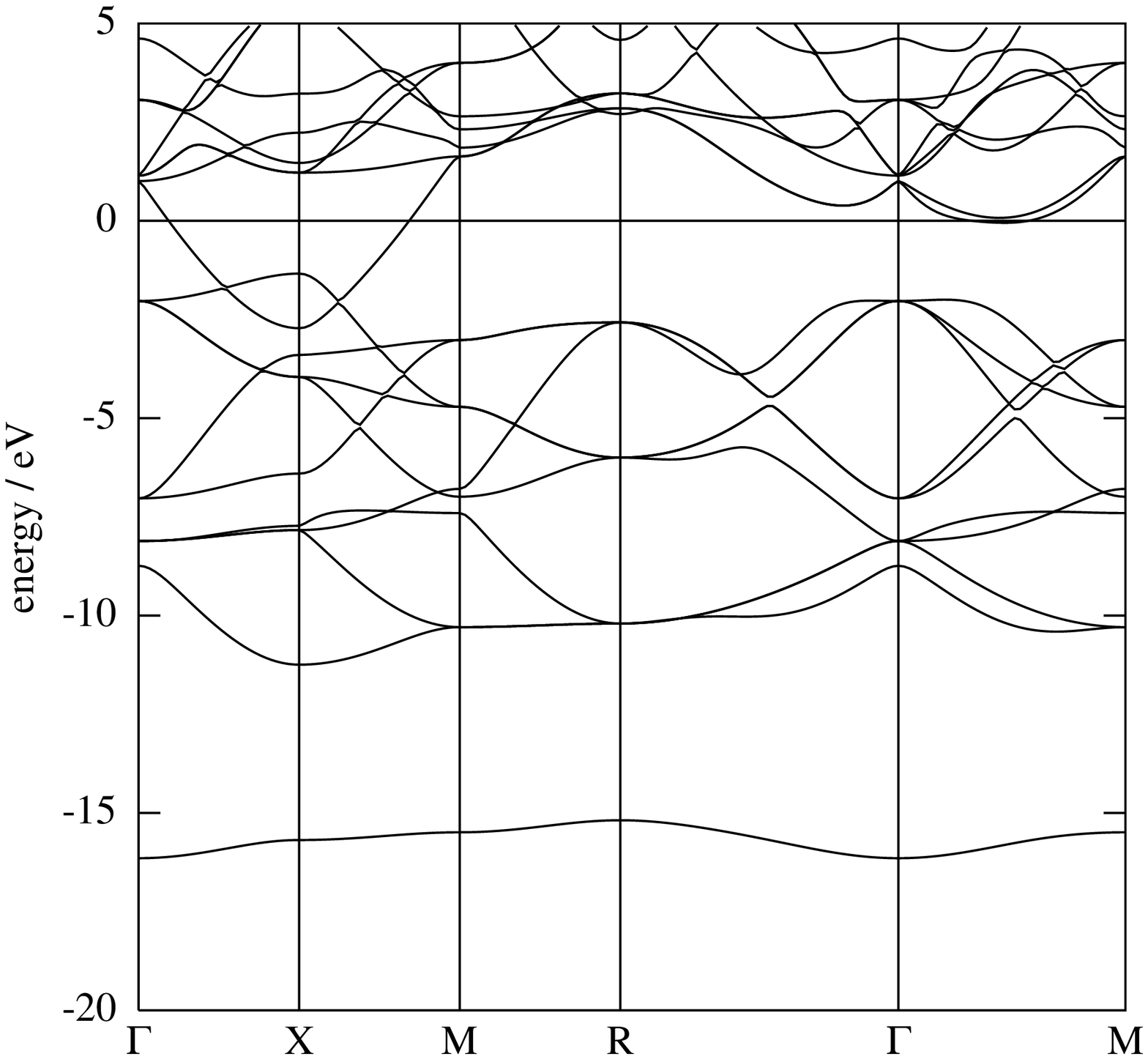}}
\end{center}
\caption{Electronic band structure for YB$_4$ (left) and YB$_6$ (right).}
\end{figure}

The DOS curves for YB$_4$ and YB$_6$ are given in Fig.~3 together with the
local partial DOS components. For both borides the states at energies
of ca.~$-15$~eV are dominated by B $2s$ character with a considerable amount
of B $2p$ character. Such states appear for all B atoms constituting the
B$_6$ octahedra, but not for B(2), which belongs to the B$_2$ unit of which each
B atom is linked to two different octahedra. In this energy range hybride
orbitals are formed which lead to strong covalent $\sigma$ bonds between
the boron atoms within a B$_6$ unit which can be seen from the corresponding
electron densities shown in Fig.~4.

After an energy gap of about 3.7~eV a complex of bands appears in both compounds
where in the lower energy range B $s$ states prevail with an admixture of
B $p$ states except for the B(2) atoms. These states correspond mainly
to bonds between different B$_6$ octahedra and $s$--$s$ $\sigma$ bonds 
in the B$_2$ units of YB$_4$. At higher energies below the Fermi level
B $p$ states are predominant apart from a relatively small B $s$
and a considerable Y $d$ contribution. Also in this energy range the
inter\-octahedral bonds are dominant except for the region very near the
Fermi level in YB$_6$ where the intraoctahedral B--B bonds again become important.

In YB$_6$ the Y $d$ states have mostly $e_g$ character which can be seen from
the valence electron densities in Fig.~5. This Figure also shows that this
is not the case for YB$_4$ although near the Fermi level $e_g$-like
character, i.~e., $(d_{z^2},d_{x^2-y^2})$ character, is predominant.

Our results for YB$_6$ are in very good agreement with the detailed
semiempirical tight-binding investigation by Longuett-Higgins and 
Roberts~\cite{lr-prsl-336-54}. In order to make the comparison we
performed crystal-field splittings for the B $p$ (and Y $d$) states.
We thus obtained the distinction between their radial ($p_z$) and tangential
orbitals ($p_x,p_y$). Together with electron-density plots for 
certain energy ranges, the regions for the different site orbitals
for the B$_6$ octahedron could be identified. We obtained the following
order (increasing energy): $a_{1g}$ (band at ca.~$-15$~eV), $a'_{1g}$,
$t_{1u}$, $e'_g$, and $t_{2g}$. (The nomenclature is taken from Ref.~\cite{lr-prsl-336-54}
and refers to the irreducible representations for the whole octahedron.) 

In Fig.~6 the valence electron densities in the (001) plane through the
B atoms in YB$_4$ and YB$_6$ is displayed. Apart from the structural differences
between both borides (irregular heptagonal arrangement of the B atoms outside
the octahedra in YB$_4$ compared to an almost regular octagon in YB$_6$)
the slightly higher valence electron densities between the B atoms outside
the octahedra in both compounds can be seen as well as the deep density
minima which designate the much weaker covalent Y--Y bonds perpendicular
to the (001) plane.

\vspace{0.5cm}
\begin{figure}[p]
\begin{center}
\mbox{\includegraphics[width=1.00\hsize]{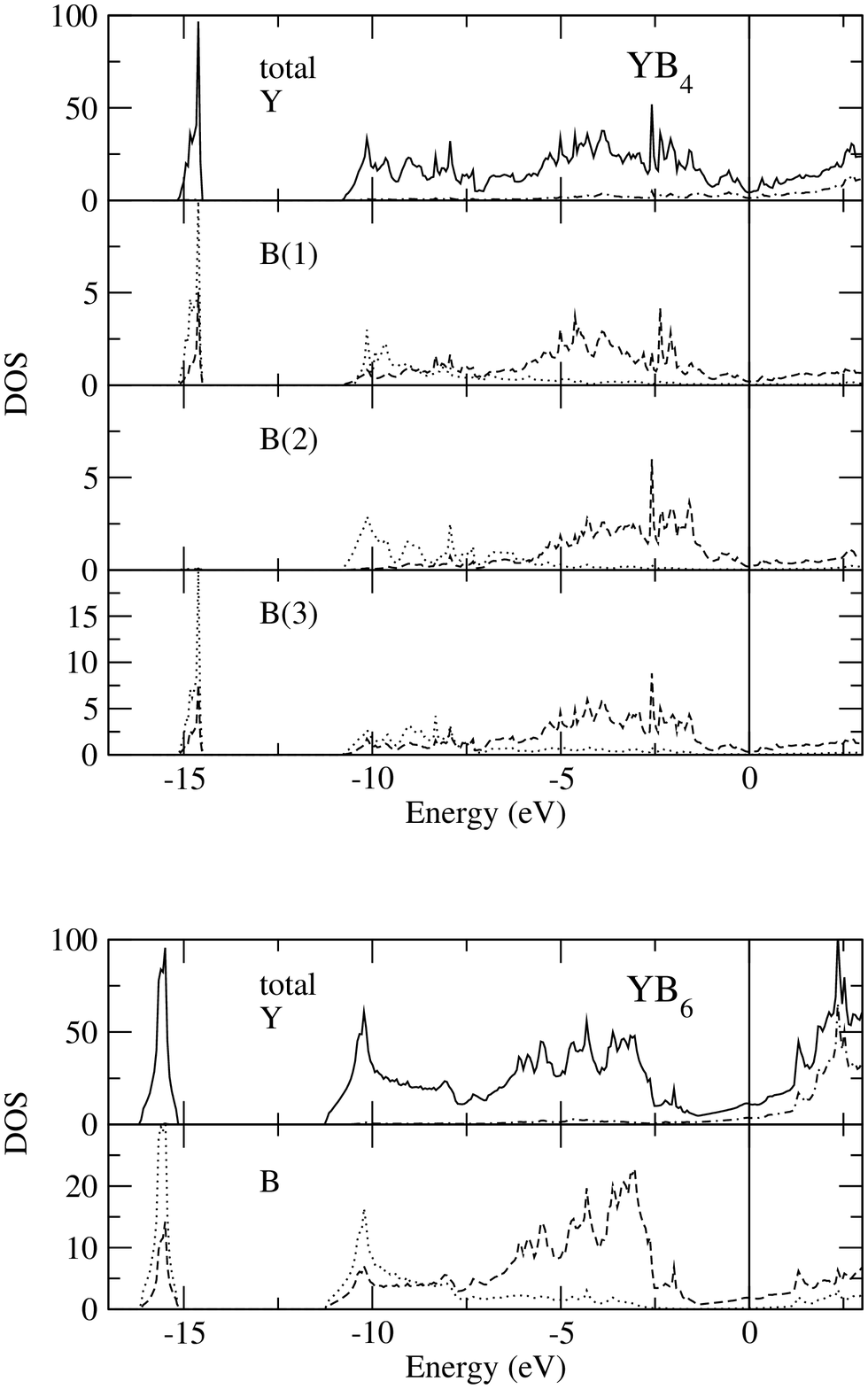}}
\end{center}
\caption{Total DOS (full line) and local partial DOS components (B $s$: dotted line;
B $p$: dashed line; Y $d$: dash-dotted line) for YB$_4$ (top) and YB$_6$ (bottom) in
units of states per Rydberg and per formula unit.}
\end{figure}

\begin{figure}
\begin{center}
\makebox[\textwidth][s]{\includegraphics[width=0.45\hsize,clip=true]{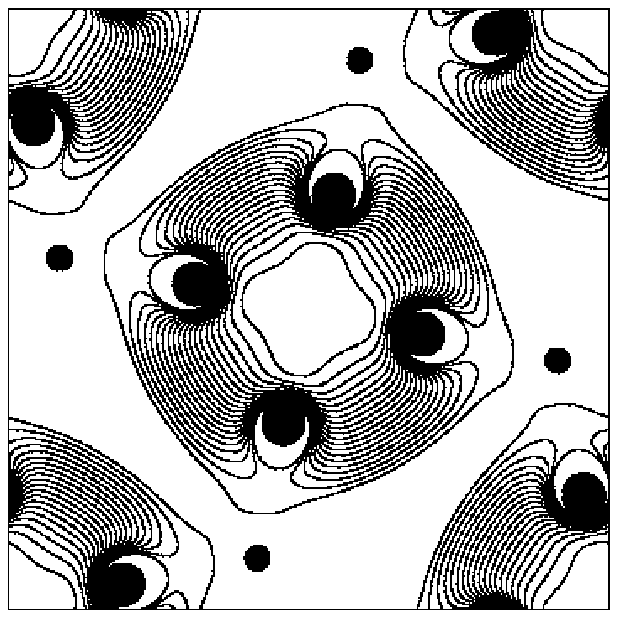}\hfil
\includegraphics[width=0.45\hsize,clip=true]{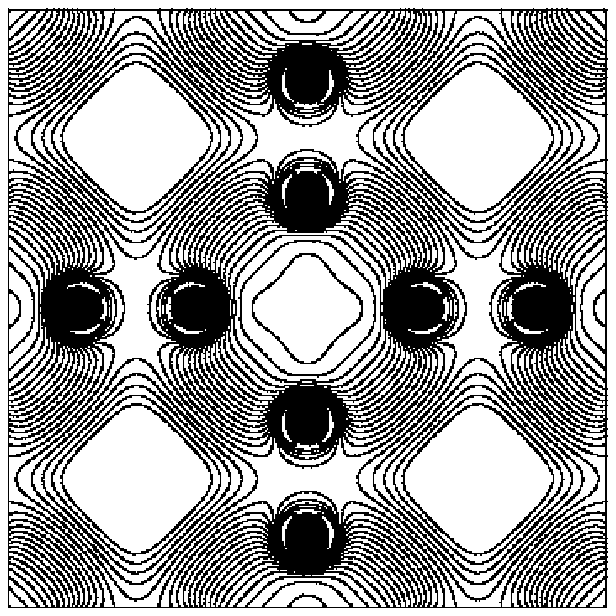}}
\end{center}
\caption{Electron densities in the (001) plane through the B atoms for
the lowest valence bands at ca.~$-15$~eV. Left: YB$_4$, right: YB$_6$.
A logarithmic grid of contour lines has been used ($x_i=x_0\,2^{i/3}$).
For YB$_4$ the B(2) atoms appear as black bullets without a surrounding 
electron density.}
\end{figure}

\begin{figure}
\begin{center}
\makebox[\textwidth][s]{\includegraphics[width=0.45\hsize,clip=true]{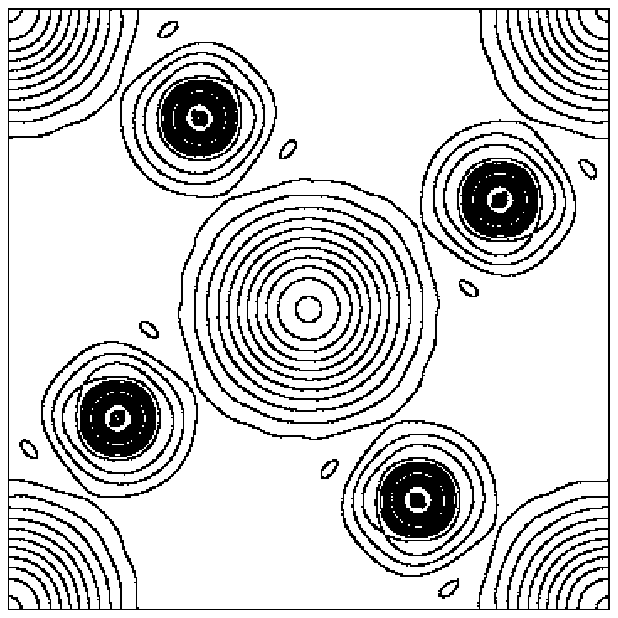}\hfil
\includegraphics[width=0.45\hsize,clip=true]{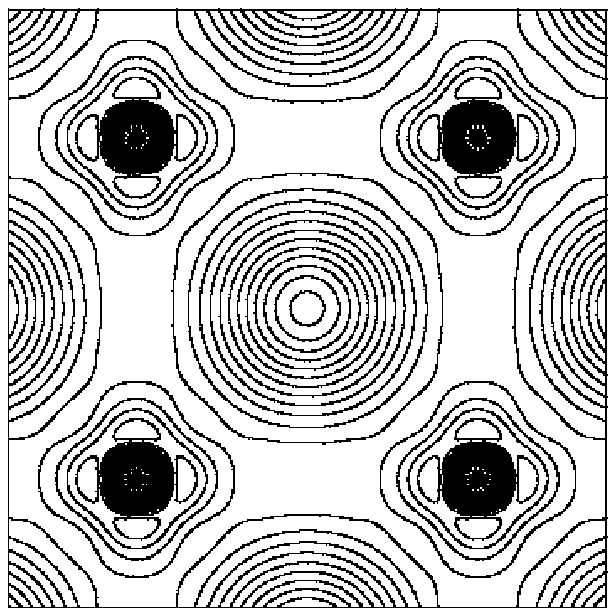}}
\end{center}
\caption{Valence electron densities in the (001) plane through the Y atoms. 
Left: YB$_4$, right: YB$_6$. A logarithmic grid of contour lines has been
used ($x_i=x_0\,2^{i/3}$).}
\end{figure}

\begin{figure}
\begin{center}
\makebox[\textwidth][s]{\includegraphics[width=0.45\hsize,clip=true]{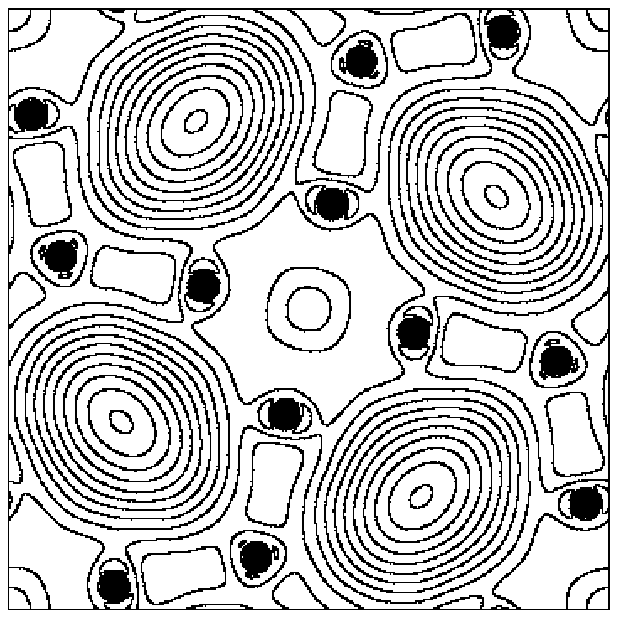}\hfil
\includegraphics[width=0.45\hsize,clip=true]{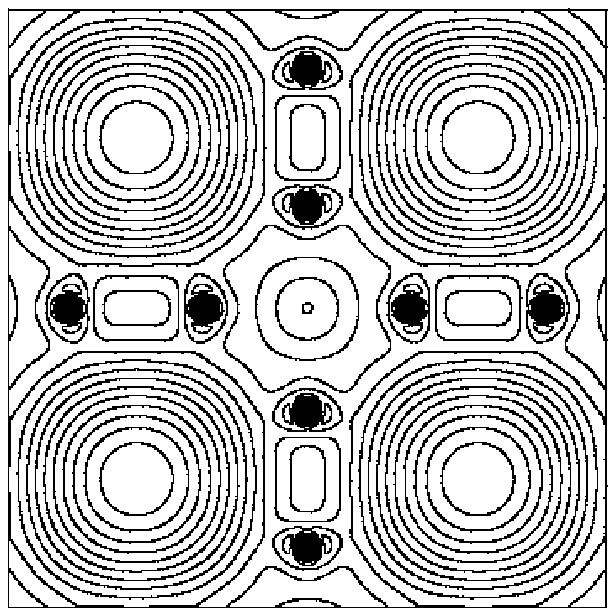}}
\end{center}
\caption{Valence electron densities in the (001) plane through the B atoms.
Left: YB$_4$, right: YB$_6$. A logarithmic grid of contour lines has been
used ($x_i=x_0\,2^{i/3}$).} 
\end{figure}

\section{Summary}
For YB$_4$ and YB$_6$ we have performed electric-field gradient and
Knight shift measurements for the B sites. The aim of the paper has been
a comparison of the NMR results and the available structural data with
the results of accurate first-principles calculations. 
We have therefore optimized the structures by atomic forces and stress-tensor
minimization. For the optimized structures EFG calculations have been performed.
Perfect agreement was found between the experimental and calculated
structural parameters and electric-field gradients thus confirming the 
structure models available in the literature. Based on calculated total
and local partial DOS as well as electron-density plots, we analyse the
bonding situation in both compounds and conclude that the lowest valence
bands are of particular importance for the stabilty of both yttrium borides.

\begin{ack}
The authors would like to thank P. Vajda for stimulating discussions and the
Austrian Science Foundation (project no.~\mbox{P15801-N02}) for financial support.
The calculations were performed on the Schr\"odinger~II Linux cluster of the
Vienna University Computer Centre.
\end{ack}

\end{document}